\documentclass[aps,prl,nofootinbib]{revtex4-1}

\usepackage{amssymb,graphics,graphicx}
\usepackage{footnote}
\usepackage{fullpage}

\begin{document}

\title
{\Large \bf Seeking Evolution of Dark Energy }

\author{Paul H. Frampton\footnote{frampton@physics.unc.edu} and 
Kevin J. Ludwick\footnote{kludwick@physics.unc.edu}}

\affiliation{Department of Physics and Astronomy, University of North Carolina,
Chapel Hill, NC 27599-3255}

\date{\today}
\begin{abstract}
\begin{center} \textbf{Abstract} \end{center}
We study how observationally to distinguish between a cosmological constant (CC)
and an evolving dark energy with equation of state $\omega(Z)$.
We focus on the value of redshift $Z^{*}$
at which the cosmic late time acceleration begins and $\ddot{a}(Z^{*}) = 0$.
Four
$\omega(Z)$ are studied, including the well-known CPL model and a new model that has advantages 
when describing the entire expansion era.
If dark energy is represented by a CC model with $\omega \equiv -1$, the present ranges for $\Omega_{\Lambda}(t_0)$ and
$\Omega_m(t_0)$ imply that $Z^{*} = 0.743$ with $4\%$ error. We discuss
the possible implications of a model-independent measurement of $Z^{*}$ 
with better accuracy.

\end{abstract}

\maketitle

\newpage

\noindent
\emph{\textbf{Introduction}} ~~~ The interface between astrophysics and
particle physics has never
been stronger than now because our knowledge of gravity comes in large part
from observational astronomy and cosmology. At particle colliders, seeking the constituent
of dark matter is an important target of opportunity.

\bigskip

\noindent
Dark energy is widely regarded as the most important issue in all of
physics and astronomy. Other than the Cosmological Constant (CC) model,
and the very interesting, if not yet fully satisfying, usage of string theory,
there is no compelling theory. So we are motivated to pursue a purely 
phenomenological approach to attempt to make progress towards the understanding of dark energy.

\bigskip

\noindent
The discovery of cosmic acceration \cite{perlmutter,kirshner} in 1998
has revolutionized theoretical cosmology. The simplest theoretical
interpretation is as a CC with constant
density and equation of state (EoS) $\omega \equiv -1$. We shall refer to
alternatives to the CC model, with $\omega(Z)$ redshift-dependent,
as evolutionary dark energy.

\bigskip

\noindent
The equations which govern cosmic history, which assume Einstein's
equations, isotropy, homogeneity (FLRW metric \cite{F,L,R,W}), and
flatness as expected from inflation, are (with $c=1$):

\begin{equation}
H(t)^2 = \left( \frac{\dot{a}}{a} \right)^2 = \left( \frac{8 \pi G}{3} \right) \rho 
\label{FL1}
\end{equation}

\noindent
and

\begin{equation}
\left( \frac{\ddot{a}}{a} \right) = - \frac{4 \pi G}{3} ( \rho + 3p) ,
\label{FL2}
\end{equation}

\bigskip

\noindent together with the continuity equation:

\bigskip

\begin{equation}
a \frac{d \rho}{da} = -3 (\rho + p).
\label{continuity}
\end{equation}

\bigskip

\noindent
In these equations, $p$ is pressure and $\rho$ is the density with components
$\rho = \rho_{\Lambda} + \rho_m + \rho_{\gamma}$.  Although, for small redshifts,
the radiation term is by far the smallest, we still include it.

\bigskip

\noindent
Using Eq. (\ref{FL2}), the CC model with $\omega \equiv -1$, and the WMAP7 values\footnote{Note that we use the WMAP7 \cite{WMAP7} value for $\Omega_m(t_0)$, unless
explicitly stated otherwise. We use $\Omega_{\gamma}(t_0) = 5 \times 10^{-5}$.} \cite{WMAP7}

\bigskip

\begin{equation}
\Omega_{\Lambda} (t_0) = 0.725 \pm 0.016 ~~~~ {\rm and} ~~~ \Omega_m(t_0) = 0.274 \pm 0.013,
\label{WMAP}
\end{equation}

\bigskip

\noindent
we find that \cite{Wang}

\bigskip

\begin{equation}
Z^{*} = \left( \frac{2 \Omega_{\Lambda}(t_0)}{\Omega_m(t_0)} \right)^{\frac{1}{3}}  - 1 ~~ 
= ~~ 0.743 \pm 0.030.
\label{ZstarCC} 
\end{equation}

\bigskip

\noindent
With evolution, Eq. (\ref{ZstarCC}) is modified. It is worth mentioning that 
$Z^*$ is a constant of Nature, like Hubble's constant, which can in principle 
be measured precisely, without reference to any theoretical model.  
We shall introduce various evolutionary models,
including the very popular CPL model 
\cite{CP,Lind} $\omega^{(CPL)}(Z)$ and a new proposal $\omega^{(new)}(Z)$ that 
is more physically motivated with respect to the whole 
expansion history\footnote{References \cite{CP,Lind} do, however, 
specify that the CPL model is to be used only for $0 \leq Z \lesssim 2$.}.  We
present figures which make predictions for $Z^{*}$, and we suggest 

slowly varying criteria which
support the use of $\omega^{(new)}(Z)$ over $\omega^{(CPL)}(Z)$.

\newpage

\noindent
\emph{\textbf{Evolutionary Dark Energy Models}} ~~~ When we consider dark energy models, we must
specify an equation of state $\omega(Z)$.  There is an infinite 
number of choices for $\omega(Z)$:  our objective in the present article 
is to suggest a sensible choice for the functional form of $\omega(Z)$, 
which can be valid for the entire extent of the present expansion era.  
Since there exists no compelling
evolutionary theory, we choose to consider models for $\omega(Z)$ each
containing two free parameters, which we designate as $\omega_0$ and 
$\omega_1$, each ornamented by a superscript which denotes the
model. To add more parameters would be premature. 
The first three are aleady in the literature, while
the fourth is, to our knowledge, new.

\bigskip

\noindent
(i) \underline{Linear model} ~~~ (lin) 

\bigskip

\noindent
The evolutionary equation of state (EEoS) is:

\bigskip

\begin{equation}
\omega^{(lin)} (Z) = \omega_0^{(lin)} + \omega_1^{(lin)} Z
\label{lin}
\end{equation}

\bigskip

\bigskip

\noindent
(ii) \underline{Chevallier-Polarski-Linder model} \cite{CP,Lind} ~~~ (CPL) 

\bigskip

\noindent
The EEoS for CPL is:

\bigskip

\begin{equation}
\omega^{(CPL)} (Z) = \omega_0^{(CPL)} + \omega_1^{(CPL)} \frac{Z}{1+Z}
\label{CPL}
\end{equation}

\bigskip

\bigskip

\noindent (iii) \underline{Shafieloo-Sahni-Starobinsky model} \cite{SSS} ~~~ (SSS) \bigskip

\noindent 
The SSS version of the EEoS is:

\bigskip

\begin{equation}
\omega^{(SSS)} (Z) = - \frac{1 + \tanh[(Z - \omega_0^{(SSS)}) \omega_1^{(SSS)}]}{2}
\label{SSS}
\end{equation}

\bigskip

\bigskip

\noindent
(iv) \underline{New proposal} ~~~ (new)

\bigskip

\noindent
Here we consider, as a novel EEoS, a simple modification
of the CPL EEoS:

\bigskip

\begin{equation}
\omega^{(new)} (Z) = \omega_0^{(new)} + \omega_1^{(new)} \frac{Z}{2+Z}
\label{new}
\end{equation}

\bigskip

\bigskip

\bigskip

\newpage

\begin{figure}[b]
\begin{center}
\fbox{\includegraphics[scale=0.8]{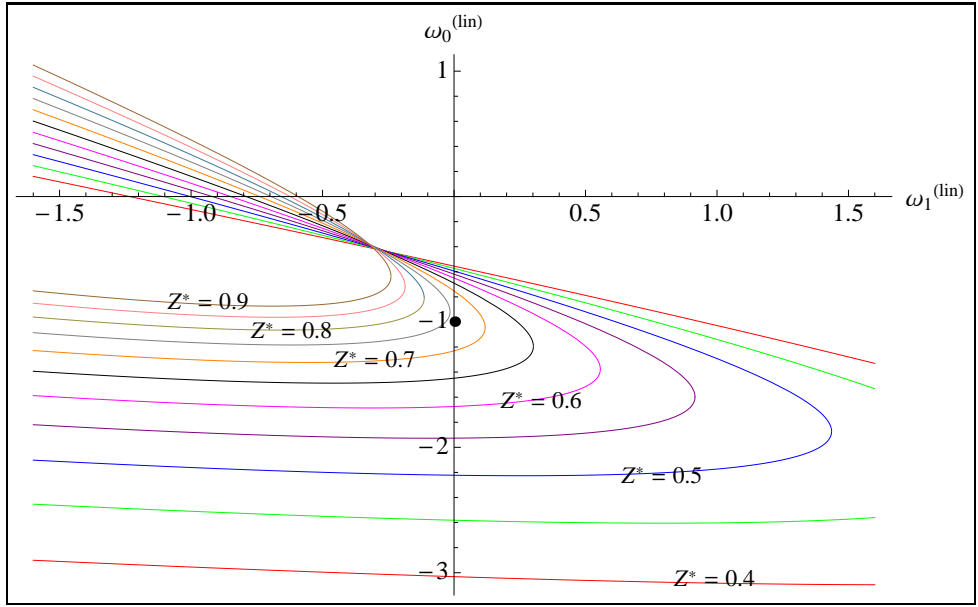}}
\caption{\footnotesize $\omega^{(lin)}_0$ is plotted against 
$\omega^{(lin)}_1$ for $0.4 \leq Z^{*} \leq 0.9$ in increments 
of $0.05$.  The dot at (0,-1) represents the CC model.  Here we 
use $\Omega_{m} (t_0) = 0.275$, which is consistent with 
the result of \cite{Virey} for the best fit for this model.}
\label{FLfig1}
\end{center}
\end{figure}

\noindent
\emph{\textbf{Analysis of the Models}} ~~~ We begin with the model where the EEoS is linear in redshift, $\omega^{(lin)}(Z)$.
In Fig. \ref{FLfig1} are shown $\omega_0^{(lin)}-\omega_1^{(lin)}$ curves for $Z^{*}$ in
the range $0.4 \leq Z^{*} \leq 0.9$. 
Several interesting features of Fig. \ref{FLfig1} deserve discussion. First, the dot
at (0, -1) confirms $Z^{*} = 0.743  \pm 0.030$ for the CC model. 
If $Z^{*}$ is measured to be $Z^{*} > 0.75$, it is necessary that $\omega_1^{(lin)} < 0$. 
If $Z^{*}$ is meaured to be $Z^{*} > 0.831$, we find that $\omega_0^{(lin)} > -1$. As $Z^{*}$ increases,
the requisite $\omega^{(lin)}(Z)$ becomes more and more distinct from the CC model.
In Fig. \ref{FLfig1}, we note that for any measured value of $Z^{*}$, 
there are two possible values of $\omega_0^{(lin)}$ for each value of $\omega_1^{(lin)}$.

\bigskip

\noindent
The EEoS for $\omega^{(lin)}(Z)$ possesses singular behavior for $Z \rightarrow \infty$
because $\omega^{(lin)}(Z) \rightarrow \pm \infty$ for $\omega_1^{(lin)}$ being positive
or negative, respectively. 

\bigskip

\noindent
The reader will remark the confluence of the $Z^{*}$-orbits in Fig. \ref{FLfig1}, which 
is an artifact of the restriction to $0.4 \leq Z^{*} \leq 0.9$.
For values of $Z^{*}$ near to but outside this range, the confluence desists. A similar
phenomenon appears in later plots.

\bigskip

\begin{figure}
\begin{center}
\fbox{\includegraphics[scale=0.8]{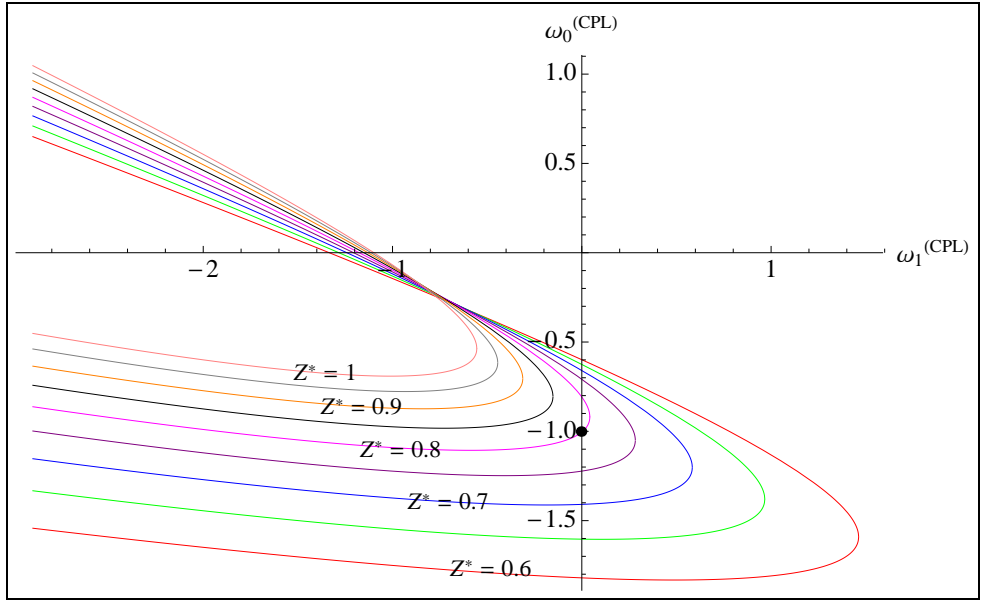}}
\caption{\footnotesize $\omega^{(CPL)}_0$ is plotted against $\omega^{(CPL)}_1$ 
for $0.6 \leq Z^{*} \leq 1.0$ in increments of $0.05$.  The dot at (0,-1) 
represents the CC model.  We use $\Omega_m (t_0) = 0.255$, which is consistent 
with the analysis in \cite{SSS} for the best fit for this model.}
\label{FLfig3}
\end{center}
\end{figure}

\begin{figure}[bp]
\begin{center}
\fbox{\includegraphics[scale=0.8]{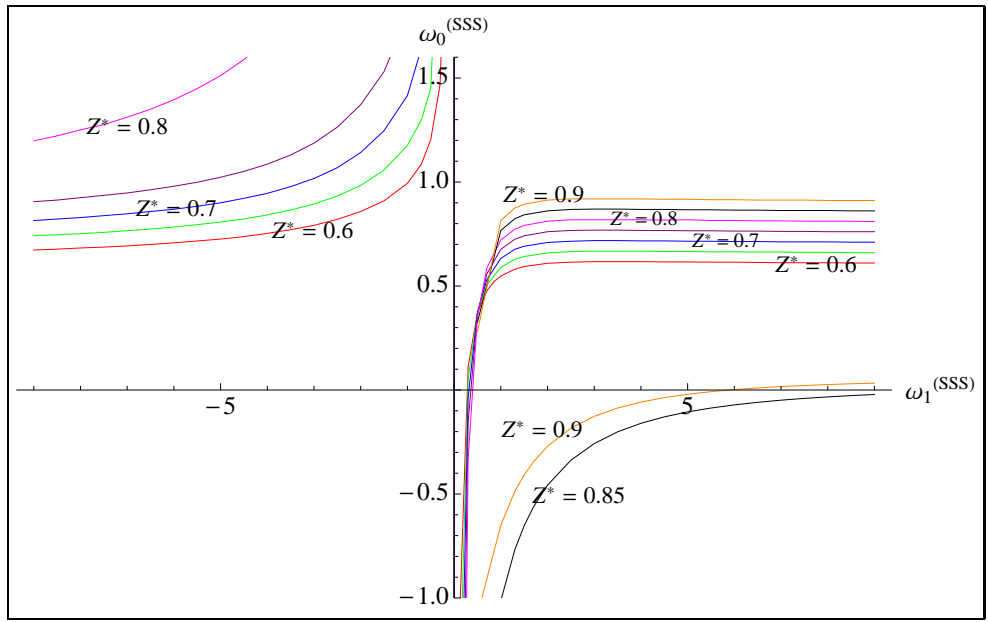}}
\caption{\footnotesize $\omega^{(SSS)}_0$ is plotted against $\omega^{(SSS)}_1$
for $0.6 \leq Z^{*} \leq 0.9$ in increments of $0.05$.
We use $\Omega_m (t_0) = 0.255$, which is consistent
with the analysis in \cite{SSS} for the best fit for this model.}
\label{FLfigSSS}
\end{center}
\end{figure}

\begin{figure}[tp]
\begin{center}
\fbox{\includegraphics[scale=0.8]{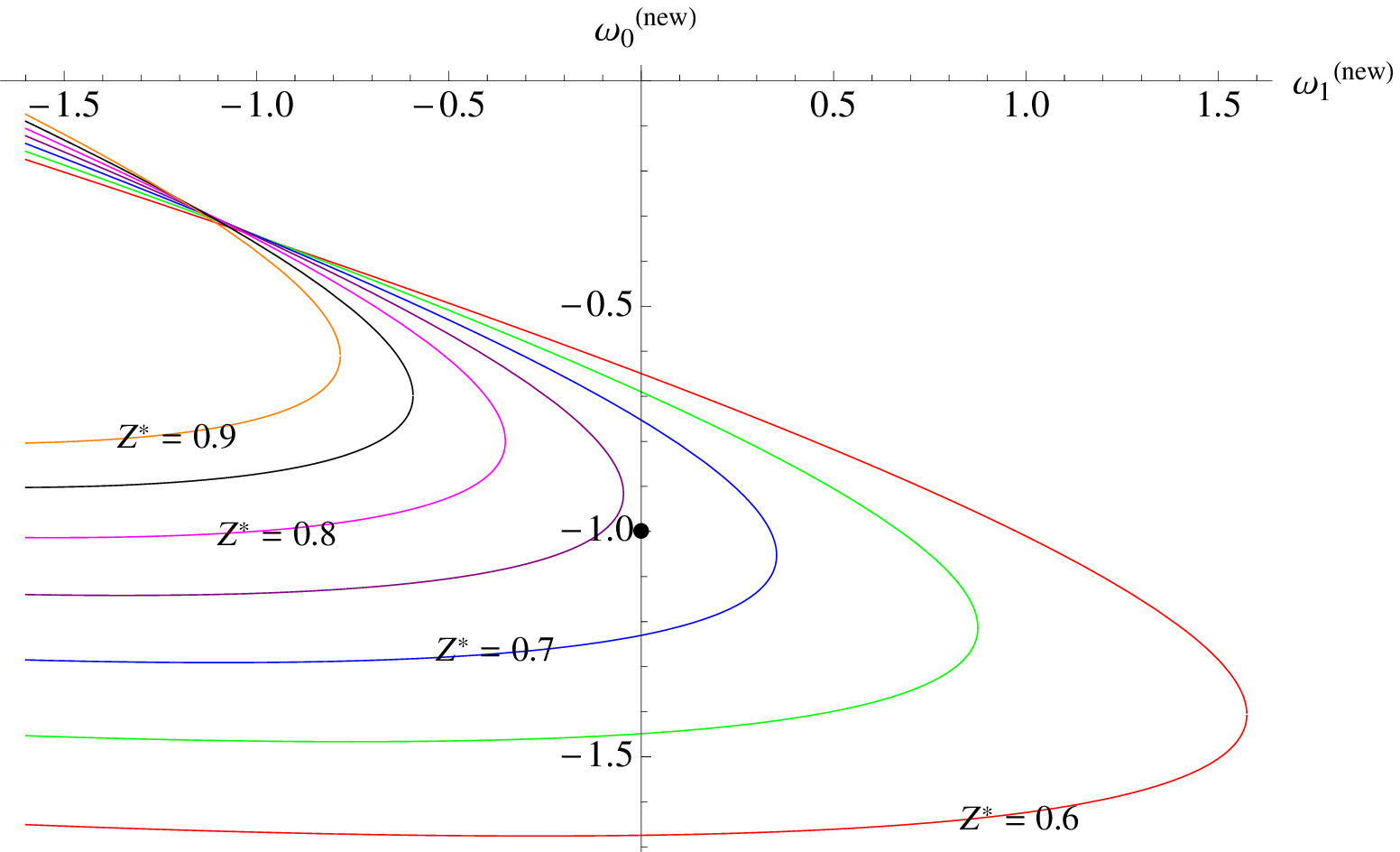}}
\caption{\footnotesize $\omega^{(new)}_0$ is plotted against $\omega^{(new)}_1$
for $0.6 \leq Z^{*} \leq 0.9$ in increments of $0.05$.  The dot at (0,-1)
represents the CC model.  We use $\Omega_m (t_0) = 0.275$.}
\label{FLfig6}
\end{center}
\end{figure}

\noindent
We next discuss the model $\omega^{(CPL)}(Z)$ \cite{CP,Lind} in which the EEoS is linear in 
$(1 - a)$, where $a$ is the scale factor.
In Fig. \ref{FLfig3} are shown $\omega_0^{(CPL)}-\omega_1^{(CPL)}$ curves for $Z^{*}$ in
the range $0.6 \leq Z^{*} \leq 1.0$. 
There are several features of Fig. \ref{FLfig3} to note. The dot
at (0, -1) confirms $Z^{*} = 0.8$, 
which is the resulting $Z^{*}$ value from Eq.(\ref{ZstarCC}) when the value $\Omega_m(t_0)=0.255$
is used, as determined by \cite{SSS} as the best fit value for the CPL model. 
If $Z^{*}$ is measured to be $Z^{*} > 0.810$, we find that $\omega_1^{(CPL)} < 0$. As $Z^{*}$ increases,
the necessary $\omega^{(CPL)}(Z)$ becomes more and more distinct from the CC model.
In Fig. \ref{FLfig3}, we see that for any measured value of $Z^{*}$, 
there are two possible values of $\omega_0^{(CPL)}$ for each value of $\omega_1^{(CPL)}$.

\bigskip
\noindent
Now we examine Fig. \ref{FLfigSSS}
for the SSS model.  We see that for a certain range of $Z^{*} > 0.8$,
which includes the best fit value of $Z^*$ which we discuss later, both
$\omega_1^{(SSS)}$ and $\omega_0^{(SSS)}$ must be greater than zero.  Note also
the degeneracy in $\omega^{(SSS)}_0$ for $Z^* = 0.85, 0.9$.

\noindent
Fig. \ref{FLfig6} displays the new model.  Once again we
see that the CC model has $Z^{*} = 0.743$.	
For $Z^{*} \geq 0.81$, $\omega_0^{(new)} > -1.0$. 
We note that for $Z^{*} > 0.75$, $\omega_1^{(new)}$ must be negative.

\bigskip

\begin{figure}[p]
\begin{center}
\fbox{\includegraphics[scale=0.8]{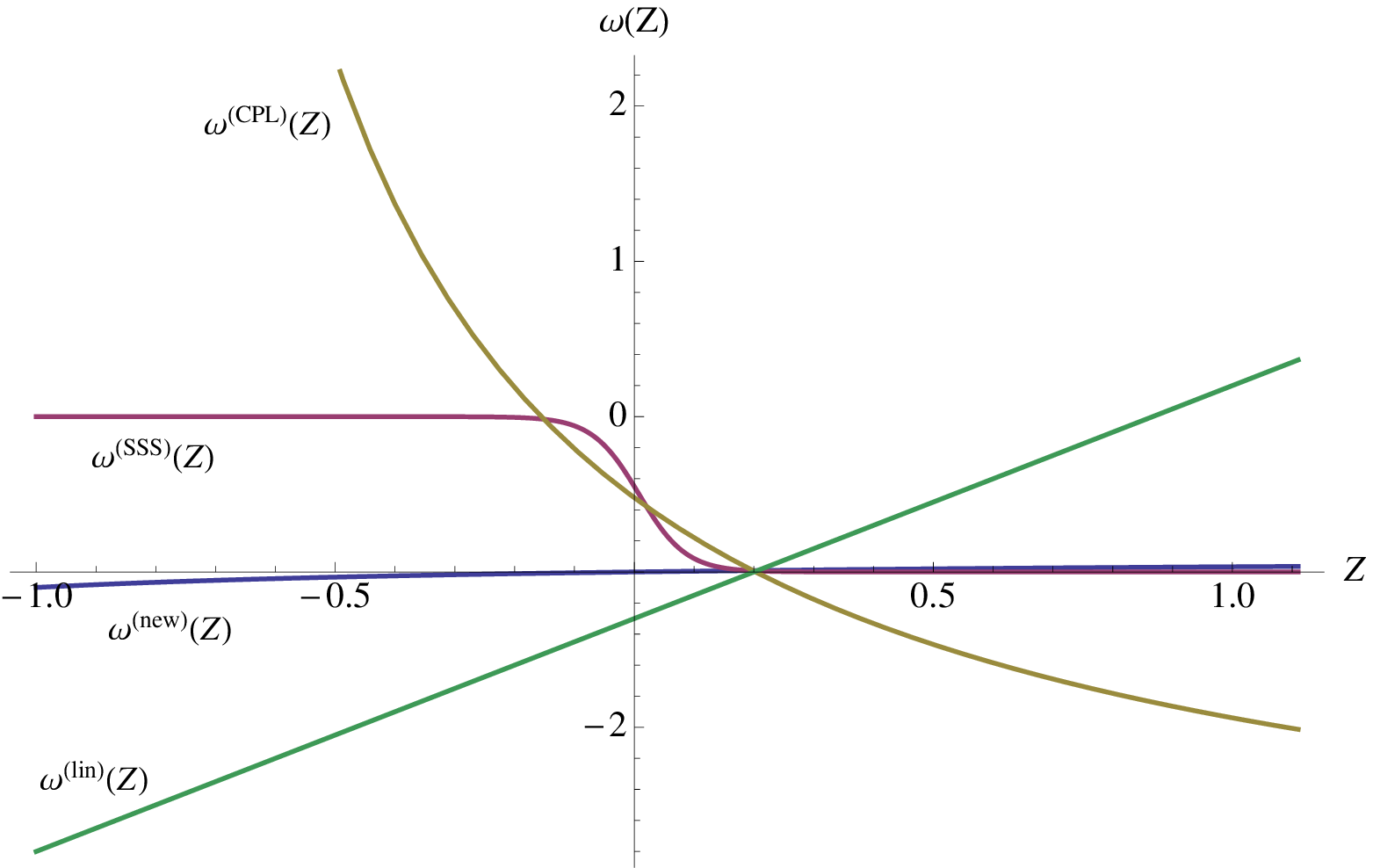}}
\caption{\footnotesize $\omega (Z)$ for all the models is plotted against $Z$ from the 
highest $Z^{*}$ value to $Z = -1$ (when the scale factor $a$ is infinite).  
The horizontal axis at $\omega (Z) = -1$ is the 
line representing the CC model.  
We use $\omega^{(lin)}_0 = -1.3$, $\omega^{(lin)}_1 = 1.5$, and
$Z^* = 0.4067$ (from the best fit for this model given in \cite{Virey}). 
We use $\omega^{(CPL)}_0 = -0.522$,
$\omega^{(CPL)}_1 = -2.835$, and $Z^* = 1.11$ for the CPL model and 
$\omega^{(SSS)}_0 = 0.008$, $\omega^{(SSS)}_1 = 12.8$, and $Z^* = 0.855$
for the SSS model (both from the respective best fits given in \cite{SSS}). 
$\omega^{(new)}(Z)$ is plotted using $\omega^{(new)}_0 = -1$ 
and $\omega^{(new)}_1 = 0.1$, which gives $Z^* = 0.729$.
}
\label{FLfig8}
\end{center}
\end{figure}

\noindent
In Fig. \ref{FLfig8}, we show $\omega^{(lin)}(Z)$, $\omega^{(CPL)}(Z)$, and $\omega^{(SSS)}(Z)$ 
as a function of $Z$ (including the future, $-1 \leq Z < 0$) 
using the best-fit parameters of \cite{Virey} for $\omega^{(lin)}(Z)$ 
and those of \cite{SSS} for the other 2 models.  $\omega^{(new)}(Z)$ is also plotted using the choice of
$\omega^{(new)}_0 = -1$ and $\omega^{(new)}_1 = 0.1$.  

\bigskip

\noindent
Unlike the CC model, where the 
future of the universe is infinite exponential expansion, 
the best fit for $\omega^{(lin)} (Z)$ necessarily
leads to a big rip, at a finite time in the future \cite{FLS}.  The EEoS for 
$\omega^{(CPL)}(Z)$ possesses singular behavior for $Z \rightarrow -1$
because $\omega^{(CPL)}(Z) \rightarrow \pm \infty$ for $\omega_1^{(CPL)}$ being negative
or positive, respectively.  For the SSS model, $\omega^{(SSS)}(Z)$ varies in the range
$0 > \omega^{(SSS)}(Z) > -1$.

\bigskip

\noindent
As for our fourth and last model $\omega^{(new)}(Z)$, from Eq.(\ref{new}), we
note that this choice has the advantage that for all $Z$ this EEoS lies between
$(\omega_0^{(new)} - \omega_1^{(new)})$ and $(\omega_0^{(new)} + \omega_1^{(new)})$. This
is illustrated in Fig \ref{FLfig8} where, for the choices
$\omega_0^{(new)} = -1.0$ and
$\omega_1^{(new)} = +0.1$, the EEoS falls smoothly from $-0.9$ at the big bang to $-1.1$
at the big rip.

\bigskip

\noindent
\emph{\textbf{Slowly Varying Criteria}} ~~~ We now discuss which $\omega(Z)$ for dark energy is
the best for observers to employ. Regretfully, there is no theoretical guidance about 
evolution. Nevertheless, we here propose slowly varying criteria which is based purely
on grounds of aesthetics and, especially, conservatism.
All the present data are consistent with the CC model $\omega^{(CC)} \equiv -1$. 
We choose to remain proximate to it, as supported by \cite{Sarkar, Serra}.
One consideration is that we prefer any global $\omega(Z)$ to have analytic, non-singular 
behavior for $Z \rightarrow -1$ and $Z \rightarrow \infty$. 
Finally, we propose to impose the inequality representing conservatism,

\begin{equation}
|\omega(Z) + 1 | \ll 1  ~~~  {\rm for} ~~ {\rm all}   ~~~  -1 \leq  Z <  \infty.
\label{slow}
\end{equation}

\bigskip

\noindent
Next, we consider application of our slow variation criteria to
the four specific models we have discussed in tbe present article.
For this task, Fig. \ref{FLfig8} will be used.  To be fair to their inventors, 
these EEoS were intended to apply for only a limited range of redshift.

\bigskip

\noindent
(i) \underline{Linear model}

\bigskip

\noindent
By studying the EEoS in Eq. (\ref{lin}), we notice that for $Z \rightarrow + \infty$, $\omega^{(lin)}(Z)$ 
approaches $\pm \infty$ depending on the sign of $\omega_1^{(lin)}$.  Also, $\omega^{(lin)}(Z)$ violates 
the criterion of Eq. (\ref{slow}).  We conclude that this linear model is disfavored according to our 
slow variation criteria.

\bigskip

\noindent
(ii) \underline{CPL model}

\bigskip

\noindent
By examining the EEoS in Eq. (\ref{CPL}), we note that as $Z \rightarrow -1$, 
$\omega^{(CPL)}(Z) \rightarrow \mp \infty$ for a ($\pm$) sign for $\omega^{(CPL)}_1$.  
Therefore, according to our criteria, this model is disfavored. 

\bigskip

\noindent
(iii) \underline{SSS model}

\bigskip

\noindent
Looking at Eq. (\ref{SSS}), we see that 
$\omega^{(SSS)}(Z)$ varies from $0$ to $-1$ for the best fit given in \cite{SSS},
so it is non-singular.  By this token, however, 
it does not satisfy Eq. (\ref{slow}).  This model, then, is also disfavored by our criteria.  
It should be noted, though, that \cite{SSS} studied this EEoS as a toy model to illustrate
the importance of an EEoS that fits data well for small and large positive $Z$.

\bigskip

\noindent
(iv) \underline{New model}

\bigskip

\noindent
This model is non-singular for all $-1 \leq Z < \infty$ because [$Z/(2+Z)$] varies smoothly from $-1$ to $+1$,
 as can be seen from examining Eq. (\ref{new}).
It also satisfies Eq. (\ref{slow}) if we choose appropriate values for $\omega^{(new)}_0$ and $\omega^{(new)}_0$. 

\bigskip

\noindent
\emph{\textbf{Discussion and Conclusions}} ~~~ The outstanding observational question about dark 
energy is whether it is a CC model with 
$\omega(Z) \equiv -1$ or an evolutionary model with a non-trivial EEoS.  The model-independent 
observational measurement of $Z^*$ is very useful for making this 
distinction.

\bigskip

\noindent
The theoretical prediction of $Z^*$ is, however, dependent on the EoS that is assumed.  The value 
for the CC model using the WMAP7 values of $\Omega_m(t_0)$ and $\Omega_{\Lambda}(t_0)$ is $Z^* = 0.743 \pm 0.030$. 
As the data become even more precise, the error on $Z^*$ will diminish, making it observationally easier 
to detect deviation, if any, from the CC model.  

\bigskip

\noindent
In the four EEoS models listed earlier, the possible values of $Z^*$ for different values of the parameters
$\omega_0$ and $\omega_1$ can be read off from our plots.  These plots show that degeneracies appear. 
For a given $Z^*$ and a specific type of EEoS, there is an allowed curve in the $\omega_1$-$\omega_0$ plane.  
For all $0.4 \leq Z^* \leq 1.0$, there exist disallowed regions in the $\omega_1$-$\omega_0$ plane. 

\bigskip

\noindent
To go further, we have to give criteria for selecting one EEoS.  The most popular choice in the last few years
has been the CPL model \cite{CP,Lind} because it approximates the linear model at low redshift.  Also, it
has a simple interpretation in terms of the scale factor:

\begin{equation}
\omega^{(CPL)}(Z) = \omega_0^{(CPL)} + \omega_1^{(CPL)} (1 - a(Z))
\label{CPLa}
\end{equation}

\noindent
However, as $a(Z) \rightarrow \infty$ ($Z \rightarrow -1$), the CPL model diverges.  Of course, the authors of 
\cite{CP,Lind} intended their model to be applicable only for a limited range of $Z$.  However, it seems preferable
for $\omega(Z)$ to cover the entire range of cosmic history, both the past and future.  

\bigskip

\noindent
Our novel EEoS also approximates the linear model at low redshift.  It has, like the CPL model, a straightforward
physical interpretation in terms of the scale factor:

\begin{equation}
\omega^{(new)}(Z) = \omega_0^{(new)} + \omega_1^{(new)} \left(\frac{1 - a(Z)}{1 + a(Z)}\right)
\label{newa}
\end{equation}

\noindent
We think one advantage of this new model over the CPL model is that it is non-singular for $-1 \leq Z < \infty$.  
A second advantage is that it can satisfy the slow variation criterion of Eq. (\ref{slow}).  

\bigskip

\noindent
In conclusion, the choice of an evolutionary alternative to the CC model depends theoretically on constraining the 
EEoS, and we have proposed a new EEoS which not only has a simple physical interpretation but also is well-behaved 
for all possible redshifts.  The model-independent extraction of $Z^*$ from  
observational data is a familiar process \cite{Daly}. A more accurate model-independent estimate of $Z^{*}$
by global fits to all relevant data is worthwhile.
It is an interesting issue how the present
considerations of evolutionary dark energy are related to the possible occurence of a big rip \cite{FLS}.  
This requires ultra-negative pressures of dark energy, so it
is interesting that situations involving phantom energy have appeared in the
context of extra spatial dimensions in string theory \cite{SW}.

\bigskip

\noindent
It has been argued \cite{PHF} that dark energy effects can be detected
only by studying physical systems as large as galaxies. Thus, it is unlikely
that any terrestrial experiment can be sensitive to dark energy.

\bigskip

\noindent
Understanding dark energy may, or may not (in the CC model), require a gravitational
theory more complete than general relativity, which has been accurately confirmed \cite{Will}
only
at the scale of the solar system, say $\sim 10^{12}$ meters, while dark energy
operates above the galactic size, say $\sim 10^{20}$ meters. Thus, it is likely,
even probable, that study of dark energy will inform us, in the
near future, how to go beyond Einstein,
which is the most important direction both for particle physics and astrophysics.  

\bigskip

\newpage

\section{acknowlegements}
\noindent
This work was supported in part by DOE Grant No. DE-FG02-05ER41418.

\end{document}